\documentclass[useAMS,usenatbib]{mnras}  
\usepackage{graphicx}
\usepackage{color}
\usepackage{graphicx}
\usepackage{txfonts}
\usepackage{color}
\usepackage{natbib}
\usepackage{epsfig}
\usepackage{hyperref}
\usepackage{longtable}
\usepackage[switch]{lineno}

\title[Broadband IDV in 3C~279]{Probing the gamma-ray variability in 3C~279 using broadband observations}
\author[B. Rani et al.]{B. Rani$^{1,2}$\thanks{NPP Fellow},
T. P.\ Krichbaum$^{2}$,
S.-S. Lee$^{3,4}$,
K. Sokolovsky$^{5,6,7}$,
S. Kang$^{3,4}$,
\newauthor
D.-Y. Byun$^{3,4}$,
D. Mosunova$^{5}$,
J. A. Zensus$^{2}$\\
        $^{1}${NASA Goddard Space Flight Center, Greenbelt, MD 20771, USA}   \\
        $^{2}${Max-Planck-Institut f{\"u}r Radioastronomie (MPIfR), Auf dem H{\"u}gel 69, D-53121 Bonn, Germany} \\
        $^{3}${Korea Astronomy and Space Science Institute, 776 Daedeok-daero, Yuseong-gu, Daejeon 34055, Korea} \\
        $^{4}${University of Science and Technology, 217 Gajeong-ro, Yuseong-gu, Daejeon 34055, Korea} \\
        $^{5}${Astro Space Center of Lebedev Physical Institute, Profsoyuznaya 84/32, 117997 Moscow, Russia} \\ 
        $^{6}${IAASARS, National Observatory of Athens, Vas.~Pavlou \& I.~Metaxa, GR-15~236 Penteli, Greece} \\
        $^{7}${Sternberg Astronomical Institute, Moscow State University, Universitetskii~pr. 13, 119992 Moscow, Russia} 
}

\begin{document}

\date{Accepted --- . Received ---}

\pagerange{\pageref{firstpage}--\pageref{lastpage}} \pubyear{2016}

\maketitle

\label{firstpage}

\begin{abstract} 
We present the results of a broadband radio-to-GeV observing campaign organized to get a better 
understanding of the radiation processes responsible for the $\gamma$-ray flares observed in 3C 279.
The total intensity and polarization observations of the source were carried out between December 28, 2013 and 
January 03, 2014 using the {\it Fermi}-LAT, {\it Swift}-XRT, {\it Swift}-UVOT, and KVN telescopes. 
A prominent flare observed in the optical/near-UV passbands  was found to be correlated with 
a concurrent $\gamma$-ray flare at a confidence level $>$95$\%$, which suggests a co-spatial origin 
of the two. Moreover, the flaring activity 
in the two regimes was accompanied by no significant spectral variations. A peak in 
the X-ray light curve coincides with the peaks of the fractional polarization curves at 43 and 86~GHz radio 
bands. No prominent variation was noticed for the total intensity and  the electric vector position angle (EVPA) 
observations at radio bands during this period. We noticed a possible hint of steepening of the radio spectrum 
with an increase in percentage polarization, which suggests that the radio polarization variations 
could be simply due to a spectral 
change. In a simple scenario, the correlated optical/$\gamma$-ray flares could be caused by the same population 
of emitting particles. The coincidence of the increase in radio polarization with the X-ray flux supports the picture 
that X-rays are produced via inverse-Compton scattering of radio photons. The 
observed  fractional variability for the $\gamma$-ray flare $\sim$0.23 does not exceed 
that in the optical regime, which is inconsistent with what  we usually observe for 3C~279; it could be due to different dependencies of the magnetic field and the external radiation field energy density 
profiles along the jet.  
\end{abstract}

\begin{keywords}
galaxies: active -- quasars: individual: 3C~279 --
             radio continuum: galaxies -- jets: galaxies -- gamma-rays -- galaxies: X-rays
\end{keywords}

\section{Introduction}

Powered by accretion onto super-massive black holes (with masses up to $\sim$10$^{10}$~M$_{\odot}$),
active galactic nuclei are extremely bright objects in the extra-galactic sky. 
In a small sub-group of these objects, a
substantial fraction of accretion energy is converted into kinetic energy forming highly
collimated and relativistic outflows of energetic plasma and magnetic fields, called jets.
Blazars with their jets pointing close to our line-of-sight are strong emitters of electromagnetic radiation 
over a range of more than 20 decades in energy. Because of relativistic beaming, these sources 
can be detected out to much larger distances than unbeamed objects.  
 BL Lacertae objects (BL Lacs) and flat spectrum radio quasars (FSRQs) are clubbed 
together and called blazars, in spite of the dissimilarity of their optical spectra -- 
FSRQs show strong broad emission lines, while BL Lacs have only weak or no emission 
lines in their optical spectra.
Blazars are notable for showing variability on a range of timescales that is often described 
as a superposition of multiple flares.
Despite several efforts to understand the broadband flaring activity of blazars, the 
exact origin of variability is still debated. 
Of particular interest is the question why for the same source sometimes we do see 
correlated behavior in different energy bands and sometimes we do not.

The flat spectrum radio quasar (FSRQ) 3C~279 \citep[at z = 0.538][]{burbidge1965}
is one of the most intensively studied objects of its class. The source has an estimated 
black hole mass in the range of (3--8)$\times$10$^8$~M$_{\odot}$ derived independently 
from the luminosity of broad optical emission lines \citep{woo2002} and from the width of 
the H$_{\beta}$ line \citep{gu2001}. Owing to its high flux density and 
prominent variations in total intensity and polarization, 3C~279 is an excellent 
candidate to examine physics of extragalactic jets and to understand particle acceleration to 
high energies.   It has been monitored intensively at radio,
optical, and more recently also X-ray and $\gamma$-ray frequencies and has been
the subject of intensive multi-wavelength campaigns \citep[e.g.,][]{maraschi1994, 
hartman1996,  chatterjee2008,  wehrle1998, larionov2008, collmar2010, bottcher2007, hayashida2012, 
hayashida2015, kang2015}.  In 2006, emission at very high energies (E$>$100~GeV) was detected from 3C~279 with 
the Major Atmospheric Gamma-Ray Imaging Cherenkov (MAGIC) telescope \citep{albert2008}.

The source has a bright radio jet extending up to kiloparsec scales. The very long baseline interferometry (VLBI) 
observations have measured apparent velocities from 4 to 20~$c$ in the parsec scale region of the 
jet, which is aligned close to the observer's line-of-sight \citep[$\leq$2$^{\circ}$,][]{lister2013, jorstad2004}. 
Polarimetric observations have detected both linearly and circularly 
polarized emission from the parsec-scale jet of 3C~279 \citep{wardle1998, taylor2000, 
homan2009}. Additionally, the observed optical radiation is also highly polarized \citep[linear polarization up to 45.5$\%$ in the U~band][]{mead1990}.
\citet{wagner2001} detected variable optical circular polarization in 3C~279 
exceeding 1$\%$. \citet{abdo2010c} reported a  
coincidence of an optical polarization angle swing with a 
bright $\gamma$-ray flare in the source, which suggests a highly ordered configuration of magnetic field during 
the emission of bright $\gamma$-ray  flares.

The source has been 
detected by the LAT (Large Area Telescope) on board the {\it Fermi Gamma-ray Space Telescope} since its launch in 2008 \citep{abdo2010a}. Being one of the brightest and most rapidly variable sources in the GeV regime, 3C~279 has been the subject 
of several recent multi-wavelength campaigns \citep{hayashida2012, hayashida2015}. Multiple  
$\gamma$-ray outbursts have been detected in the source \citep{hayashida2015, paliya2015}. In December 
2013, a series of $\gamma$-ray flares were observed,
 reaching the  highest flux level measured 
in this object since the beginning of the {\it Fermi} mission, with F (E $>$100~MeV) of 
10$^{-5}$ photons cm$^{-2}$ s$^{-1}$ and a flux-doubling time scale as short as 
2~hr \citep{hayashida2015}. In June 2015, 3C~279 was observed in an exceptionally bright state 
\citep{cutini2015} with the highest measured flux, F (E $>$100~MeV) of 3.9$\times$10$^{-5}$ 
photons cm$^{-2}$ s$^{-1}$ \citep{paliya2015}, breaking its own record.

Extremely bright flashes of light in the high-energy (GeV/TeV) regime at minutes to hours timescales 
in blazars have attracted the attention of astronomers, as this suggests 
that particles can be promptly accelerated with very high efficiency in tiny magnetized emission 
regions. The rapid variations in the high-energy regime are not always 
accompanied by flaring activity at other  frequencies \citep{rani2013,rani2015, abdo2010c, hayashida2012}, making 
it even more difficult to understand 
the radiation  processes and acceleration mechanisms involved.

In the following, we present the results of our study of 3C~279 from a multi-wavelength campaign 
organized in December 2013,
when it went through a series of rapid $\gamma$-ray flares. We monitored the source using both 
ground- and space-based telescopes for a time period between December 28, 2013 and January 03, 2014. The 
aim of our study is to understand 
the origin of the observed flaring activity.
More specifically, we investigate the correlation of the $\gamma$-ray activity with the emission 
at lower frequencies.

The paper is structured as follows. Section 2 provides a brief description of observations and data 
reduction. In Section 3, we report our results. Finally, discussion and conclusions are given 
in Section 4.

\section{Multifrequency data: observations and data reduction}
From December 28, 2013, to January 03, 2014, the broadband flaring activity of the FSRQ 
3C~279 was extensively covered using both ground- and space-based observing facilities. 
The following sub-sections summarize the observations and data reduction.

\subsection{Radio observations} 
\begin{figure}
\includegraphics[scale=0.540,angle=0, trim=2 44 0 2, clip]{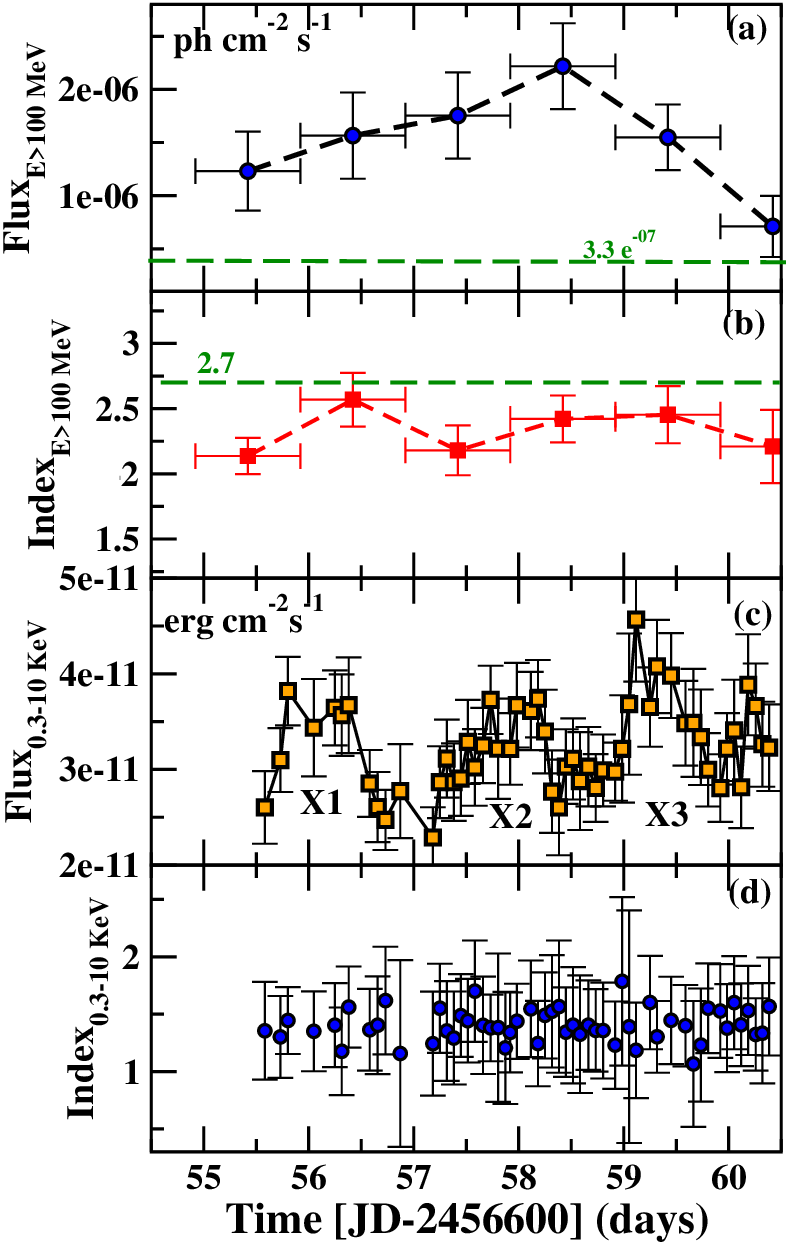}
\includegraphics[scale=0.540,angle=0, trim=-19 1 0 2.1, clip]{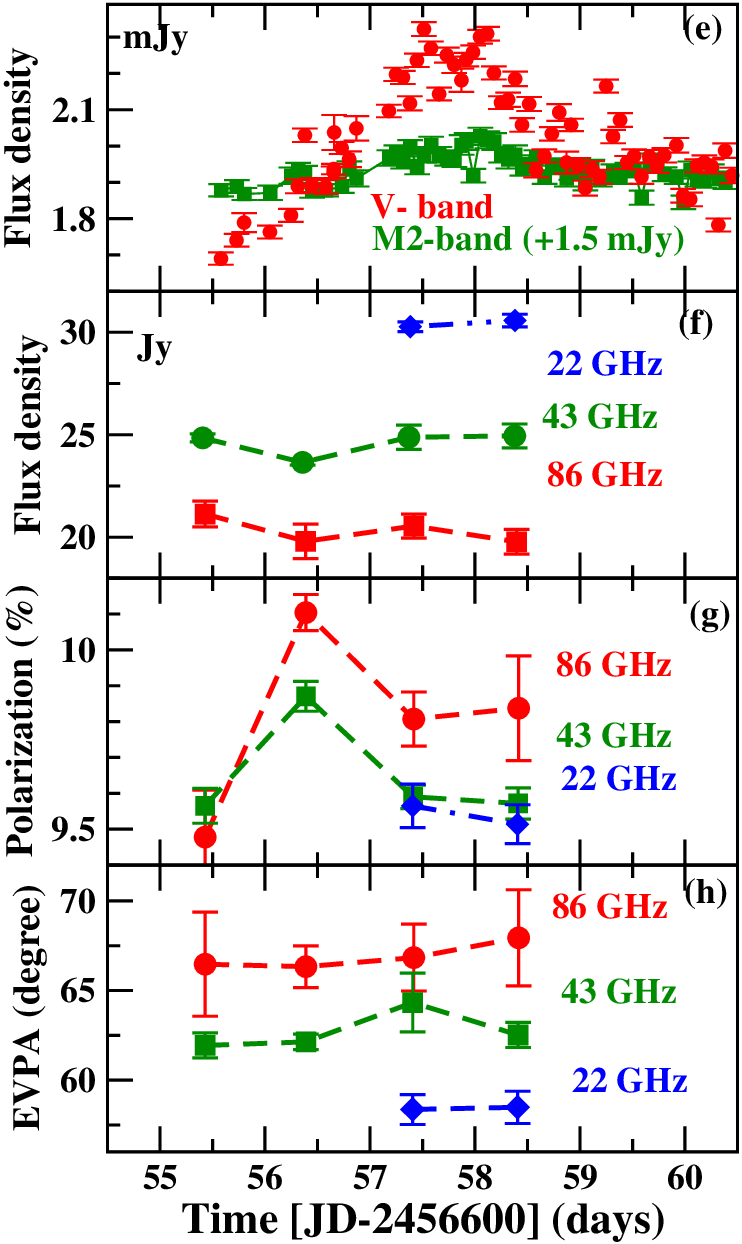}
   \caption{ Multi-wavelength observations of 3C~279: 
$\gamma$-ray photon flux (a) and photon index (b) light curves for E $>$ 100 MeV  (the dashed green 
line marks the quiescent  photon flux and index level); 
X-ray flux (c) and index (d) light curves for 0.3-10 KeV; optical V and near-UV M2 passband 
(with a 1.5 mJy offset) flux density light curves (e); 
radio flux density (f), polarization (g), and EVPA (h) light curves at 22, 43, and 86 GHz bands.          }
\label{plot_fig1}
\end{figure}

Single-dish flux density and polarization observations of the source were 
performed using the radio telescopes at individual KVN 
\citep[Korean VLBI Network\footnote{http://kvn.kasi.re.kr}][]{lee2011, lee2014} 
observatories. Observations at 22 GHz were carried out 
quasi-simultaneously using KVN Tamna, at 43 GHz using KVN Tamna and KVN Yonsei, and at 
86 GHz using KVN Yonsei. 
Polarimetric observations of the source were performed via on-off switching observations
with the on-source integration time of $\sim$6 minutes. Cross-scan pointing and antenna 
gain calibration measurements were conducted before every polarimetric observation. 
We obtained wide-band (512 MHz band-width) spectral measurements with the real and
imaginary parts of their cross-correlation  for the polarimetric observations. The
channel spacing was 125 kHz for 4096 spectral channels. The degree of linear 
polarization and the electric vector position angle (EVPA) of the source were estimated 
using a data reduction pipeline presented in \citet{kang2015}, and the rms uncertainties of the linear polarization 
observations were about 10 mJy and 15 mJy at 22 GHz and 43 GHz, respectively. The statistical 
uncertainties of the polarization angle measurements were $\leq$0.2$^{\circ}$ and $\leq$0.3$^{\circ}$ 
at 22 GHz and 43 GHz, respectively. The systematic error of the polarization angle measurements is 
2$^{\circ}$ at both frequencies. 
The rms uncertainty of the linear polarization observation is about 30 mJy at 86 GHz, and the 
statistical uncertainty of the polarization angle measurements is $<$0.5$^{\circ}$ at 86 GHz. The 
systematic error of 2$^{\circ}$ of the polarization angle measurements is also applicable to the 
86 GHz data.
Details about the data reduction pipeline for the polarimetric observations and the polarimetric 
capability of the KVN are given by \citet{kang2015}. The observed radio 
flux and polarization curves are shown in Fig.~\ref{plot_fig1} (f to h).

 \begin{figure}
\includegraphics[scale=0.35,angle=0, trim=0 0 0 0.5, clip]{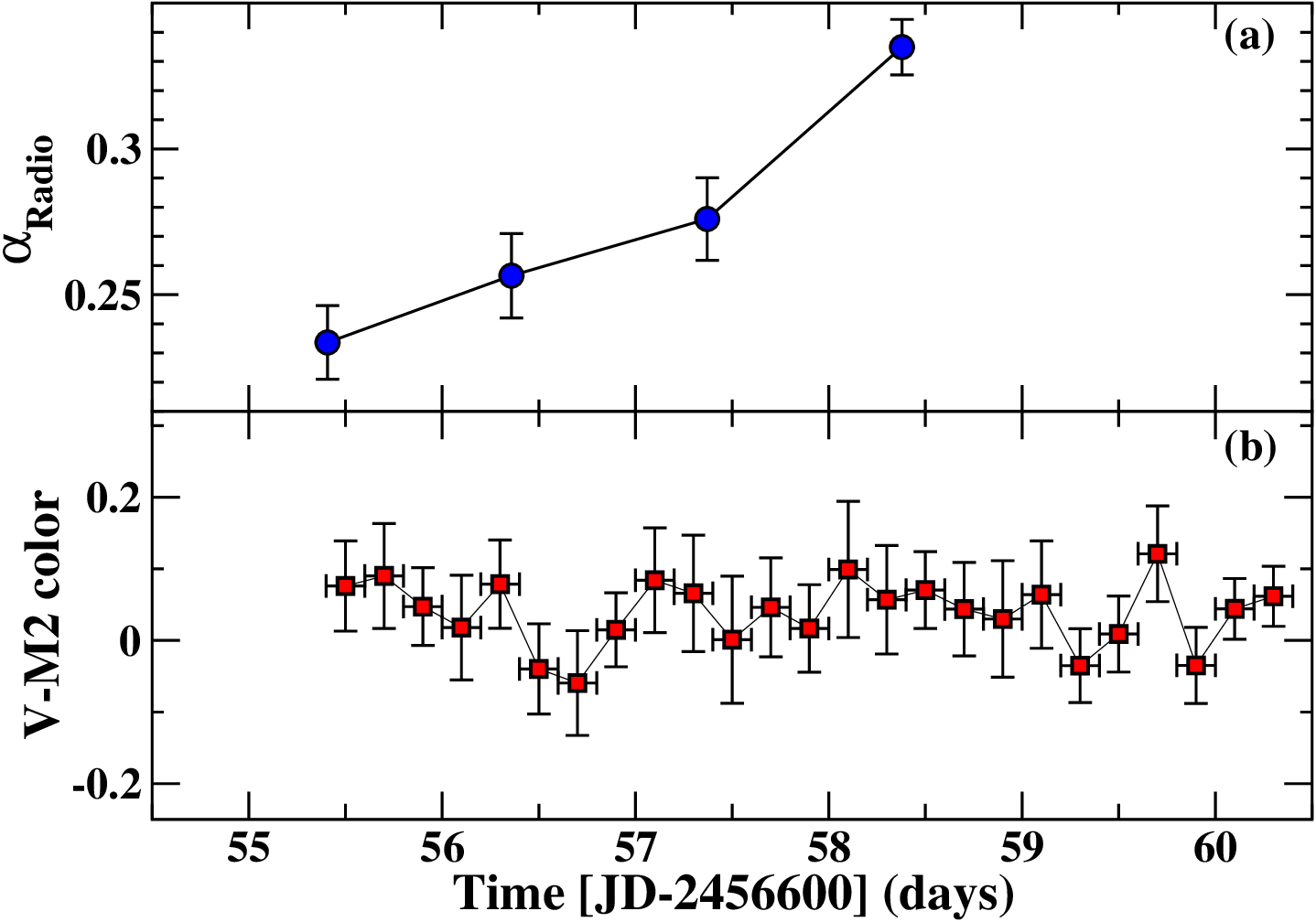}
   \caption{Top: Radio spectral index variations. Bottom: Optical V-M2 color index variations. }
\label{plot_fig1b}
\end{figure}

\subsection{Optical/UV observations}
{\it Swift}-UVOT \citep{roming2005} observed 3C~279 in optical V and
near-ultraviolet M2 filters during our campaign period. Multi-extension FITS files 
containing all images of the target obtained on a specific
date were split into single-exposure image files to preserve time resolution. All images 
were visually inspected to discard the few affected by tracking errors.
We used the {\scshape uvotsource} tool from HEASoft~v6.16 to extract the source counts from a
$5^{\prime\prime}$-radius aperture and convert them to magnitudes on the Vega scale
\citep{poole2008}. The background was estimated in an aperture annulus centered on the source, 
 with an inner radius of $10^{\prime\prime}$ and outer radius of $23^{\prime\prime}$ chosen 
to include just the  background, avoiding nearby objects.

As a consistency check, we repeated the aperture photometry measurements in
V band using the VaST code\footnote{http://scan.sai.msu.ru/vast/} that relies on 
SExtractor\footnote{http://www.astromatic.net/software/sextractor} \citep{bertin1996} for source detection and photometry.
The magnitude scale was set with comparison stars suggested for ground-based
observations\footnote{http://www.lsw.uni-heidelberg.de/projects/extragalactic/charts/1253-055}.
While the individual measurements obtained with the two methods agree within
the errors, on average the measurements obtained using the second
method are 3.5\% brighter. This difference is dominated by uncertainty in
reconstructing the non-linear detector response using the
small number of comparison stars in the field. The magnitude to flux conversion was done using the 
central wavelengths for each filter as calibrated by \citet{poole2008}. 
The observed optical  and near-UV light curves  extracted using the {\scshape uvotsource} tool
are presented in Fig.~\ref{plot_fig1}~(e).

\subsection{X-ray observations}
We analyzed 54 {\it Swift}-XRT \citep{burrows2005} observations of 3C~279 obtained in the photon counting mode simultaneously with 
{\it Swift}-UVOT. After pre-processing with {\scshape xrtpipeline}  v0.13.2, the 
spectra were extracted from a 20\,pixel-radius aperture centered on the object position, grouped to 
contain at least 25 counts in each energy bin and modeled with XSPEC~v12.9.0.
The absorbed power law with the total HI column density fixed to the Galactic
value \citep[$N_{HI} = 0.212\times10^{21}$,][]{kalberla2005} provides an acceptable ($p=0.01$) fit to 
the 0.3 - 10\,keV spectrum at all epochs.
 The X-ray photon index shows no significant changes over the course of our observations 
Fig.\ \ref{plot_fig1} (d).  The X-ray 
flux (Fig.\ \ref{plot_fig1} c) is therefore extracted keeping the photon index value fixed to 1.46.

\subsection{Gamma-ray observations}
At GeV energies, the source was observed in survey mode by the {\it Fermi}-LAT 
\citep{atwood2009}. We employed here the 100~MeV -- 300~GeV data of the source 
from December 28, 2013, to January 03, 2014.  The LAT data were analyzed using the 
standard ScienceTools (software version v10.r0.p5) and instrument response functions 
P8R2$\_$SOURCE$\_$V6. Photons in the source event class were selected for the analysis. 
We analyzed a  region of interest (ROI) of 20$^{\circ}$  radius centered at the position
of 3C~279 using a maximum-likelihood algorithm \citep{mattox1996}. In the unbinned likelihood 
analysis\footnote{http://fermi.gsfc.nasa.gov/ssc/data/analysis/scitools/likelihood\_tutorial.html}, 
we included all sources of the 3FGL catalog \citep{3fgl_paper} within 20$^{\circ}$ and the 
recommended Galactic diffuse background ($gll\_iem\_v06.fits$) and the isotropic background 
($iso\_P8R2\_SOURCE\_V6\_v06.txt$) emission components \citep{acero2016}. 
Model parameters for the sources within 5$^{\circ}$ of 
the center of the ROI were kept free.  Beyond the 5$^{\circ}$ ROI, model parameters 
of the sources reported as being significantly variable (variability index $\geq$72.44) in 
the 3FGL catalog  
are kept free as well, while the parameters for the rest were fixed to their catalog values. 
To characterize the variability properties of the source, we generated the light 
curves using a time binning of one day. The daily binned light curves of the
source at E$>$100~MeV were produced by modeling the spectra over each bin by a simple power law
(N(E) = N$_0$ E$^{-\Gamma}$, N$_0$ : prefactor, and $\Gamma$ : power law index).  Variability 
at shorter timescales cannot be investigated as the source is not bright enough. The photon 
flux (a) and photon index (b) curves are shown in Fig.~\ref{plot_fig1}. We noticed a prominent flare in the 
photon flux light curve accompanied by no clear spectral variation; the average $\gamma$-ray photon index 
is 2.4$\pm$0.2.


\section{Results}

 \begin{figure}
\includegraphics[scale=0.35,angle=0, trim=0 0 0 0.5, clip]{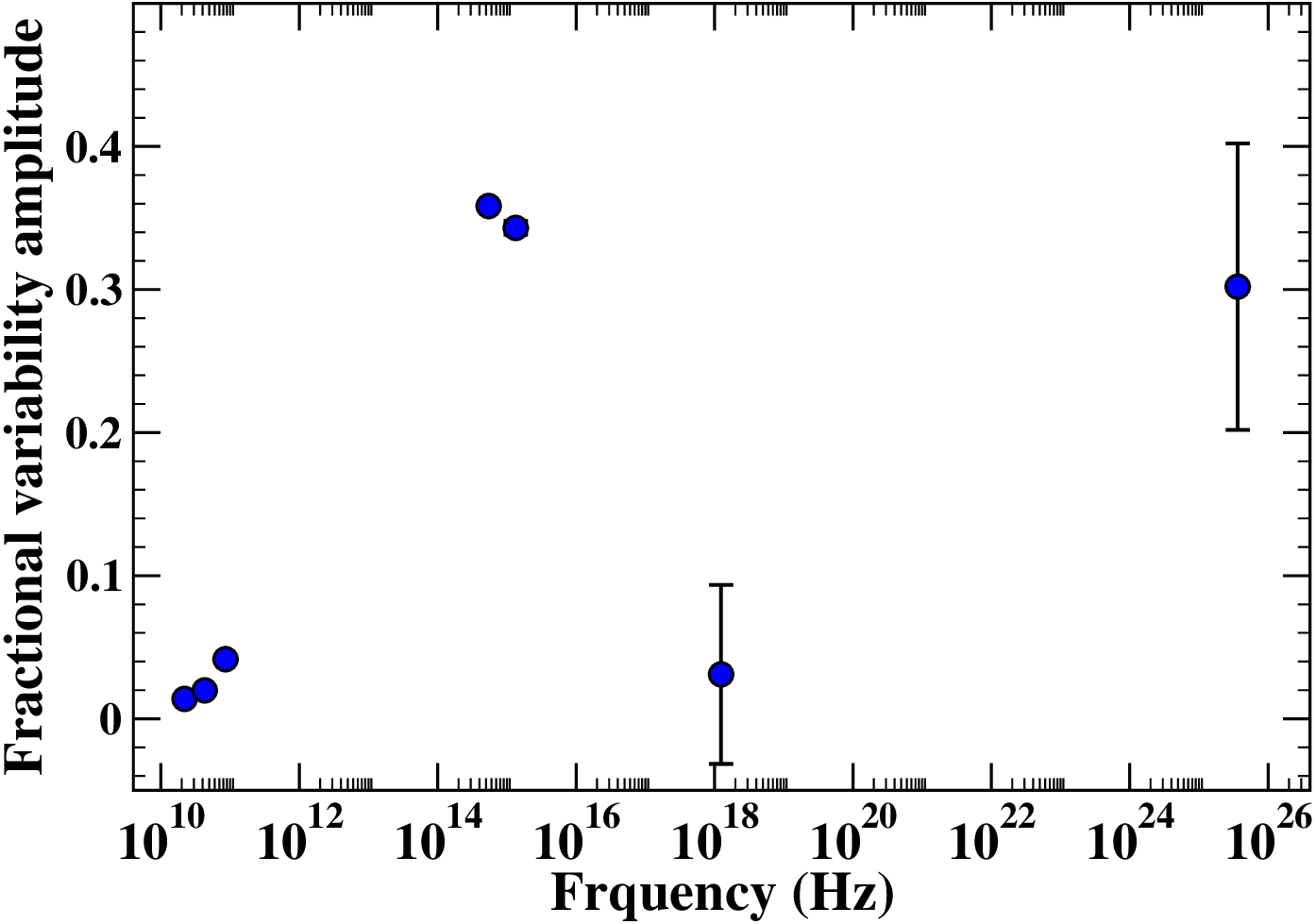}
   \caption{ $F_{var}$ amplitude spectrum of 3C~279 from radio to
$\gamma$-rays.           }
\label{plot_fig2}
\end{figure}

\subsection{Light curve analysis}
\label{sect_lc}

Figure~\ref{plot_fig1} shows the broadband light curves of the source observed between December 28, 2013, and 
January 03, 2014. At GeV energies, we observed an increase in the brightness of the source with a 
peak at JD'\footnote{JD' = JD-2456600} $\sim$58.5 (a);  the horizontal green line marks the quiescent 
flux level\footnote{The quiescent level is the photon flux value in the beginning of the outburst (November 2013).}. 
 The significance of variability is examined 
via a $\chi^2$-test (testing the hypothesis of a non-variable source). Variations in the $\gamma$-ray light curve 
are found to be marginally significant with a significance level of 2.4~$\sigma$; however, no significant 
variation (significance level $<$0.04~$\sigma$) is noticed in the $\gamma$-ray photon index curve (b). 

 We noticed three peaks (labeled as X1 to X3) at X-ray energies. The first peak (X1) roughly coincides with a 
peak in the percentage polarization curve at radio bands. The peak X2 has an apparent coincidence with the peak in 
the optical/UV light curves. The last peak (X3) is observed about a day after the peak in the $\gamma$-ray 
light curve. 
 Given the low significance of the entire X-ray light curve ($\sim$1.4~$\sigma$), we regard these flares as a possible hint of variability. 

A prominent flare was observed  in the source at optical (V band) and ultra-violet (M2 band) frequencies. The source 
brightened by a factor of two with a peak at  JD' $\sim$58 (e), which approximately coincides with the peak 
of the $\gamma$-ray flare.   The estimated significance of variability is 5.7~$\sigma$ and 4.9~$\sigma$ 
respectively for the V and M2 band light curves.
In Fig.\ \ref{plot_fig1b} (b), we show the V$-$M2 color variations during the course of our observations. 
The V-M2 color characterizing the spectral slope in the optical-UV region has an average value of 
${\rm V}-{\rm M2} = 0.05\pm0.04$ and shows no statistically significant ($<$0.1~$\sigma$) changes over the 
course of the observing campaign.

 \begin{figure*}
\includegraphics[scale=0.70,angle=0, trim=2.5 0 0 0.5, clip]{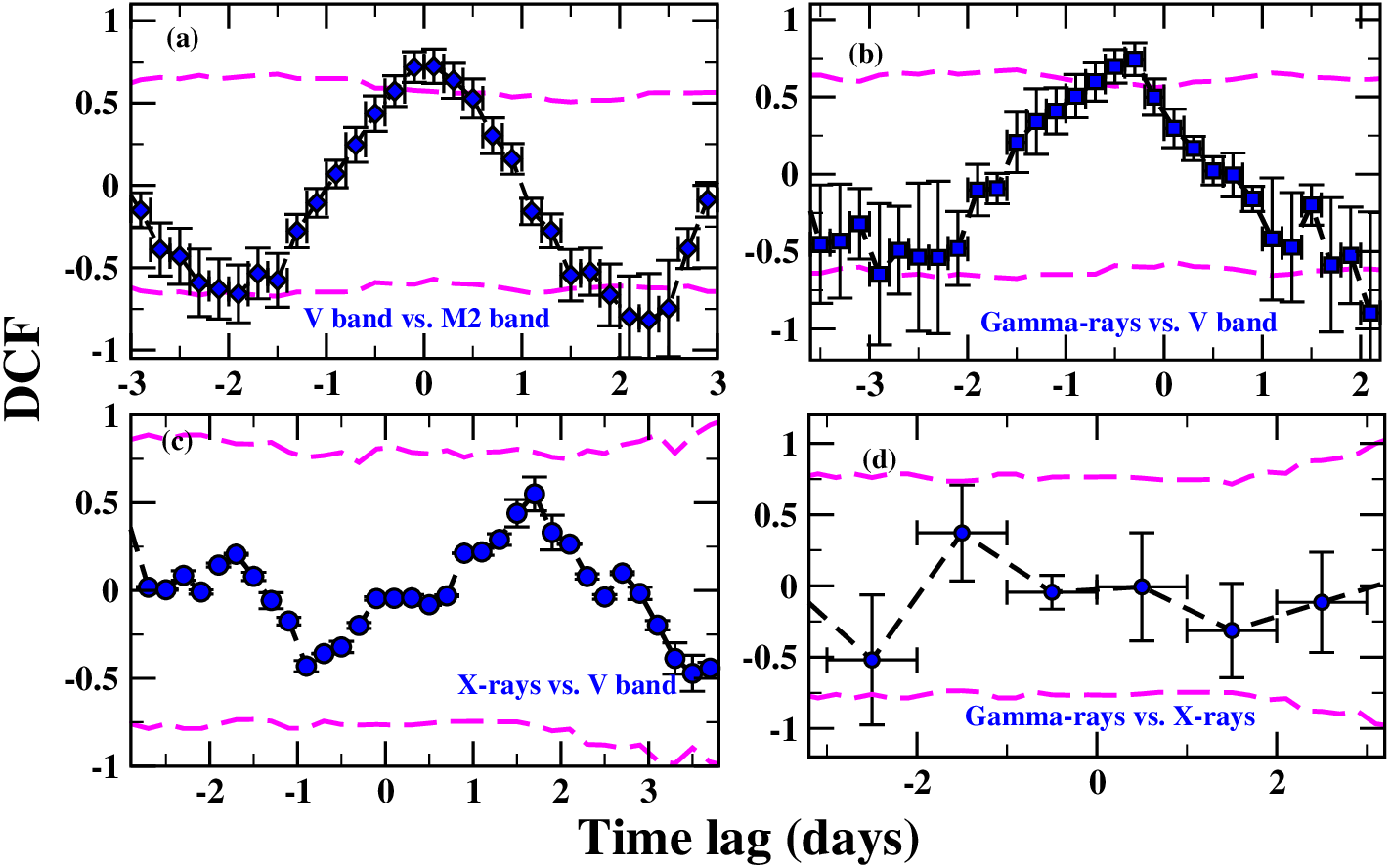}
   \caption{Cross-correlation analysis curves:  (a) Optical V band vs.\ near-UV M2 band (bin size = 0.2~day), 
(b) Gamma-rays vs. Optical V band (bin size = 0.2~day), (c) X-rays vs. Optical V band (bin size = 0.2~day), 
and (d) Gamma-rays vs. X-rays (bin size = 1.0~day).  The dashed lines denote the 95$\%$ confidence levels.     }
\label{plot_fig3}
\end{figure*}

At radio frequencies, total flux and polarization of the source were 
observed at 22, 43, and 86~GHz. 
Radio flux density (f), linear polarization (g), and EVPA (h) curves of the source are displayed in 
Fig.~\ref{plot_fig1}. The source exhibits a marginal flux density variability at radio bands; however, 
variations are more pronounced in fractional polarization curves. The observed peak in the fractional 
polarization curve at 86 and 43~GHz coincides with  the first peak, X1, in the X-ray light curve. 
Similar to total intensity, the EVPA values remain unchanged during our observations.  
The total intensity  observations at 43 and 86~GHz radio bands are used to investigate the spectral index variations. 
We used a power law of the form $P(\nu) \propto \nu^{-\alpha}$, where $\alpha$ is the spectral index, to 
get an estimate of radio spectral index. The radio spectral index variations are shown in 
Fig.\ \ref{plot_fig1b} (a). During the four days of our observations, we noticed a clear 
steepening of the radio spectrum ($\alpha$ changes from 0.23$\pm$0.02 to 0.33$\pm$0.01).

As we discussed above, the variability behavior is very complex and depends on the frequency observed. 
In the following sections, we investigate the nature of broadband flux variability and 
the possible correlation among the multi-frequency light curves of the source.

\subsection{Fractional variability}
\label{frac_var}
In order to quantify and compare the total variability at 
each observed band, we estimated the fractional variability amplitude, $F_{var}$, given by 
\citet{vaughan2003}: 
\begin{equation}
F_{\rm{var}} = \sqrt{ \frac{S^{2} -
\overline{\sigma_{\rm{err}}^{2}}}{\bar{x}^{2}}}
\end{equation}
where $S^2$ is the sample variance of the light curve, $\overline{x}$ is the average flux and   
$\overline{\sigma_{\rm{err}}^{2}}$ is the mean of the squared measurement uncertainties. 
The uncertainties on $F_{var}$ due to the measurement-error fluctuations have been estimated through 
Monte Carlo simulations by \citet[][eq.\ B2]{vaughan2003}.

The estimated $F_{var}$ is 
plotted in Fig.~\ref{plot_fig2}. 
The $F_{var}$ strongly 
depends on frequency. 
In the radio regime, F$_{var}$ increases with frequency/energy  and the trend continues until the optical/UV 
regime where it has a peak; an explanation for this is that the evolution of synchrotron flares might 
propagate from higher to lower energies \citep{marscher1985}.  The F$_{var}$ peak is followed by a dip in the X-ray domain. 
The F$_{var}$ vs.\ frequency plot (Fig.~\ref{plot_fig2})  resembles  a double-hump structure, which 
has been noted previously for several blazars \citep{chidiac2016, aleksic2015, soldi2008}. 
The double hump could be related to the two humps in the broadband 
SED (spectral energy distribution) of blazars, where the variability increases with frequency in each hump i.e.\   
largest variations are seen for the highest energy electrons producing each hump.

\subsection{Cross-correlation analysis}
\label{sect_cor}
The discrete cross-correlation function \citep[DCF,][]{edelson1988} method was used to  quantify the correlation 
among the multi-frequency light curves of the source and to search for possible time lags. The details
of this method are also discussed by \citet{rani2009}.   As a conservative approach, we used half the duration of the observations, i.e.\ $\pm$3.5~days, as the correlation search time window. Depending 
on the sampling rate of the data used for the DCF analysis, a suitable bin size was chosen  for
each observing band (see Fig.\ \ref{plot_fig3} caption).
The significance of the DCF analysis was tested 
using simulations as described in Appendix A of \citet{rani2014}.
To estimate the confidence level for each DCF 
curve, we generated a series of 1000 light curves to obtain the upper and lower bands of the 95$\%$ confidence levels.
The cross-correlation analysis results of the broadband light curves are shown in Fig.~\ref{plot_fig3}.
The dashed lines in Fig.~\ref{plot_fig3}  represent the 95$\%$\footnote{Correlations having confidence 
levels above 95$\%$ are considered significant.} confidence levels.  
In the following sections, we discuss in detail the cross-correlation analysis results.

  \begin{figure}
 \includegraphics[scale=0.32,angle=0, trim=0.5 0 0 0.5, clip]{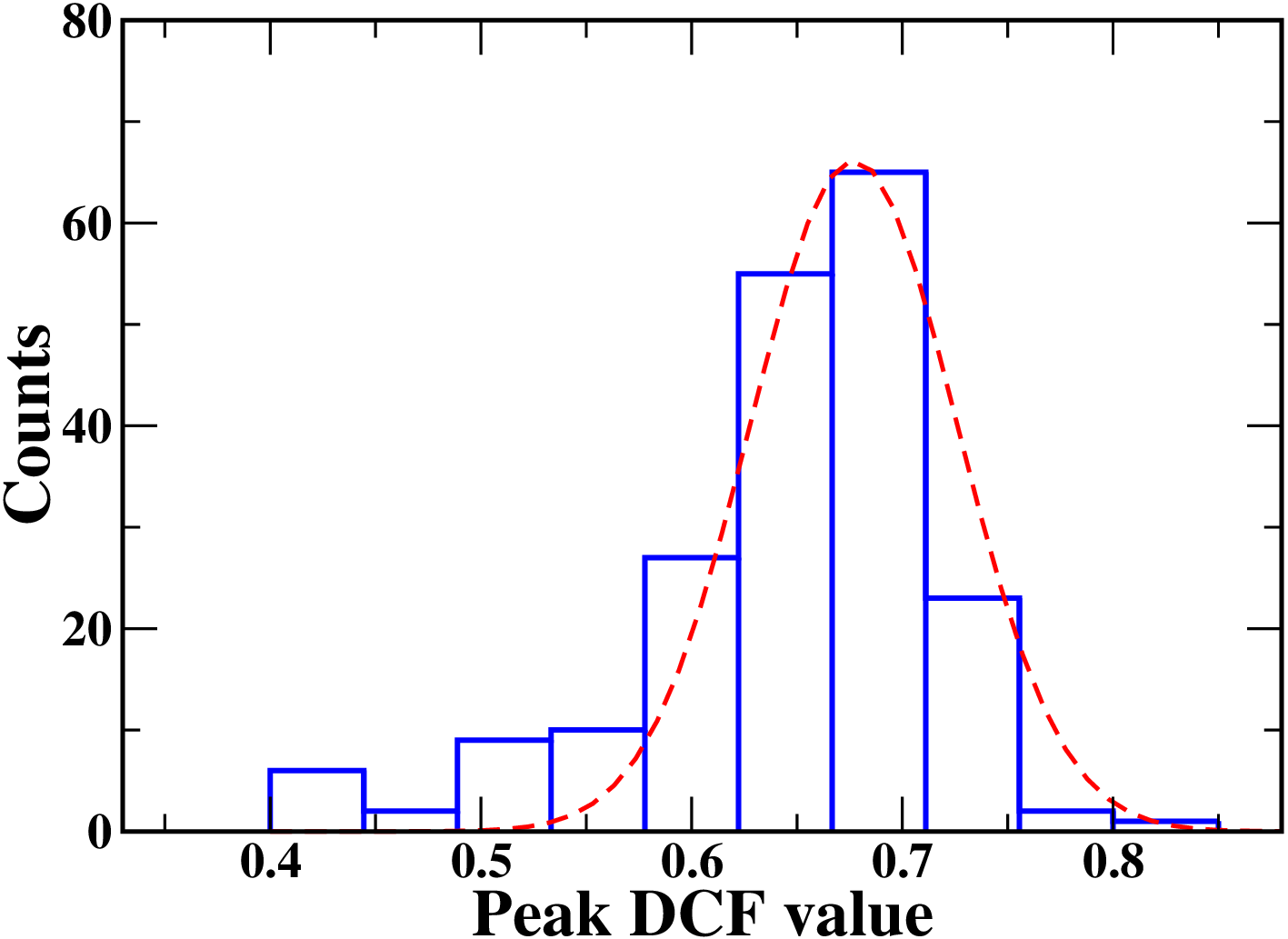}
 \includegraphics[scale=0.32,angle=0, trim=0.5 0 0 0.5, clip]{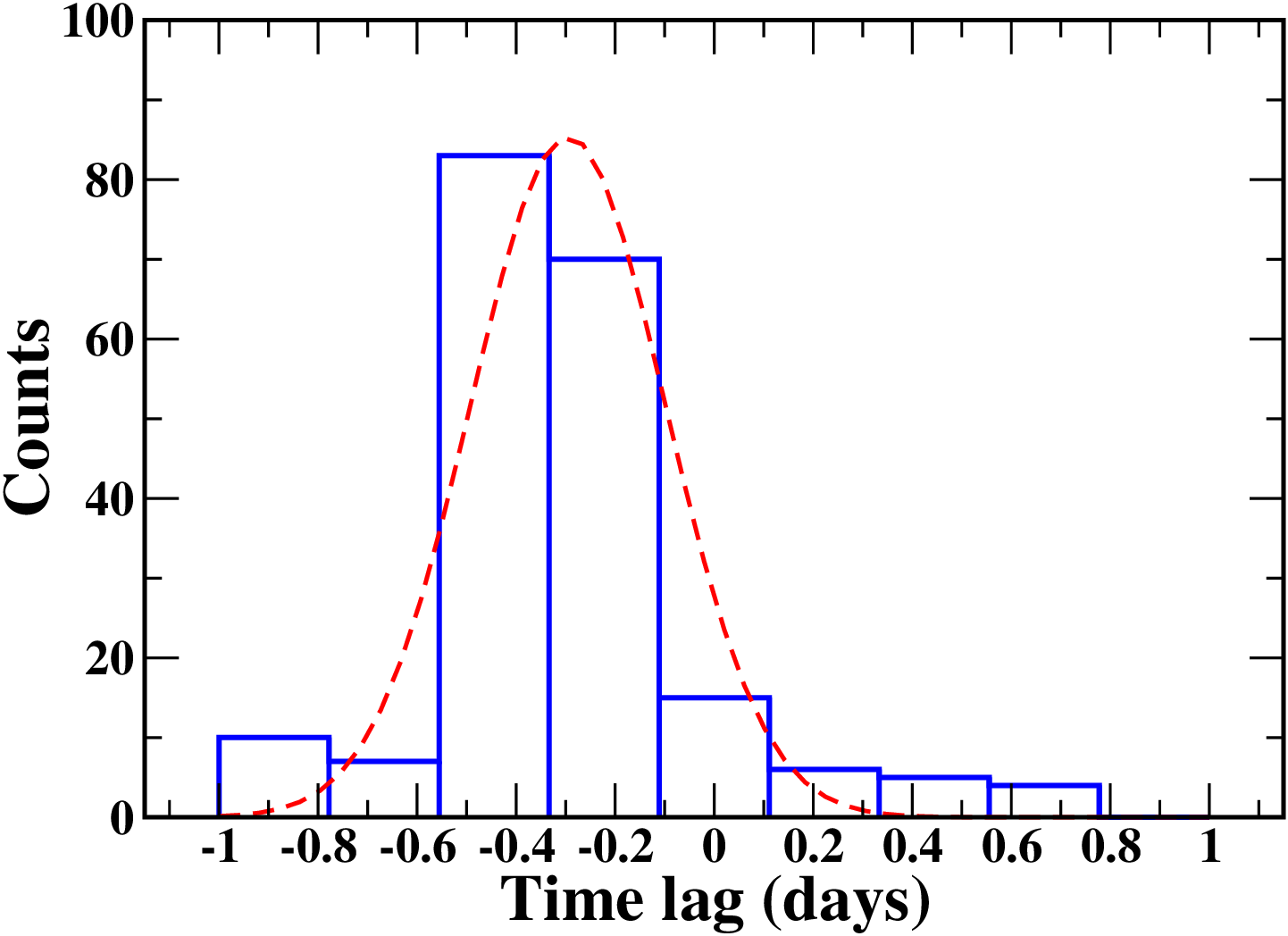}
  \caption{Histogram of peak DCF value (top) and time lag (bottom). The dashed-curve 
represents the fitted Gaussian function.   }
\label{plot_fig4a}
\end{figure}

\subsubsection{Optical vs.\ near-UV} 
A prominent flare  is observed almost simultaneously in the 
optical (V passband) and near-UV (M2 passband) light curves, which is expected as the two passbands 
are very close in frequency. A significant correlation is observed between the flux 
variations at the two bands with a correlation coefficient 0.65$\pm$0.10 at a time lag of 
0.1$\pm$0.1~days\footnote{The given uncertainty in the time lag value here and in the following sections 
is the half the bin size used for the DCF analysis.}, which is compatible with no time lag 
(Fig.~\ref{plot_fig3}~b).   
The DCF analysis therefore indicates that the flux variations 
in optical and near-UV bands are significantly correlated with  no time lag longer than 0.2~day. 
Owing to its better time sampling, we will only use the V passband data for the DCF analysis with 
broadband light curves.

\subsubsection{Gamma-ray vs.\ optical} 
\label{dcf_gamma_opt}
Visual inspection of the variability curves in Fig.~\ref{plot_fig1} shows an apparent 
similarity between the $\gamma$-ray and optical flux variations. The DCF curve between the two shows a peak 
at  $-$(0.3$\pm$0.1)~days with a correlation  coefficient 0.74$\pm$0.10 (Fig.~\ref{plot_fig3}~c) at a 
confidence level $>$95$\%$. The analysis therefore suggests that the flux variations at optical and 
$\gamma$-ray frequencies are significantly correlated with the former leading the latter by (0.3$\pm$0.1)~days.

The estimated time lag ($\sim$0.3~day) between the optical and $\gamma$-ray flares is smaller 
than   the time sampling of the $\gamma$-ray light curve (1~day).  To determine how sensitive the 
estimated time lag is to the sampling 
rate of the given light curves, we used the interpolated cross correlation 
function (ICCF) method \citep{gaskell1987, breedt2010}.   It is important to note that the simulations 
and interpolation take into account the observed profile of the light curve. If there are faster variations 
present but are not observed by any means (either due to low brightness or some other reasons), this method 
is not suitable for testing that.
We implemented the ICCF method as follows. 
We interpolated  each light curve and cross-correlated it with the other to test whether the 
sampling of the first one gives rise to spurious correlations. In the first case, we 
generated 100 interpolated $\gamma$-ray light curves having a time sampling from 0.02 to 2~days with 
a step-size of 0.02~day. The interpolated $\gamma$-ray light curves were then correlated with the observed 
optical light curve. We repeated the same procedure for the optical light curve and correlated the 
interpolated optical light curves with the observed $\gamma$-ray light curve. In total, we had 200 
DCF curves that were used to estimate the peak DCF value and the corresponding time lag. 
Figure~\ref{plot_fig4a} shows the distribution\footnote{The number of cells for the histograms was calculated 
using the ``Scott" algorithm available under the ``hist" function in $R$ (a language and environment for 
statistical computing), which computes a histogram of 
the given data. Details are given at   https://stat.ethz.ch/R-manual/R-devel/library/graphics/html/hist.html.} of 
the obtained DCF parameters (peak DCF values and time lags). 
We next investigated the distribution of the obtained DCF parameters to test how sensitive the estimated time 
lag ($-$0.3~day) is to the sampling of the observed light curves. The DCF parameter's distribution can be 
well approximated via a Gaussian function, as is shown in Fig.~\ref{plot_fig4a}. The estimated DCF value 
distribution has  a peak at 0.65 with FWHM = 0.12. For the estimated time lags, the Gaussian function yields 
a  peak at $-$0.24 with FWHM = 0.14~days. It is therefore evident that the estimated time lag is not 
sensitive to the given sampling of the light curves.

\begin{table}
\center
\caption{Fitted parameters of flares}
\begin{tabular}{lcccc} \hline
Flare                         & T$_r$         &  T$_d$        & t$_0$         & F$_0$                \\
observed at                   & (day)         &  (day)        & (JD$^{\prime}$)           &   \\\hline   
Gamma-ray                     & 2.5$\pm$1.2  &2.3$\pm$1.1   &57.9$\pm$1.2  &2.45$\pm$0.30      \\  
Optical (V)                   & 0.4$\pm$0.1  &1.8$\pm$0.4   &57.1$\pm$0.3  &1.94$\pm$0.31       \\ 
near-UV (M2)                  & 0.4$\pm$0.1  &1.0$\pm$0.3   &57.4$\pm$0.4  &0.44$\pm$0.12      \\\hline  
\end{tabular} \\
Notes: F$_0$ is in the units of mJy for optical and near-UV bands and is in 10$^{-6}$ ph~cm$^{-2}$~s$^{-1}$ 
for $\gamma$-rays. 
\label{tab_flare} 
\end{table}

\begin{figure}
   \centering
 \includegraphics[scale=0.55,angle=0, trim=0.5 0 0 0.5, clip]{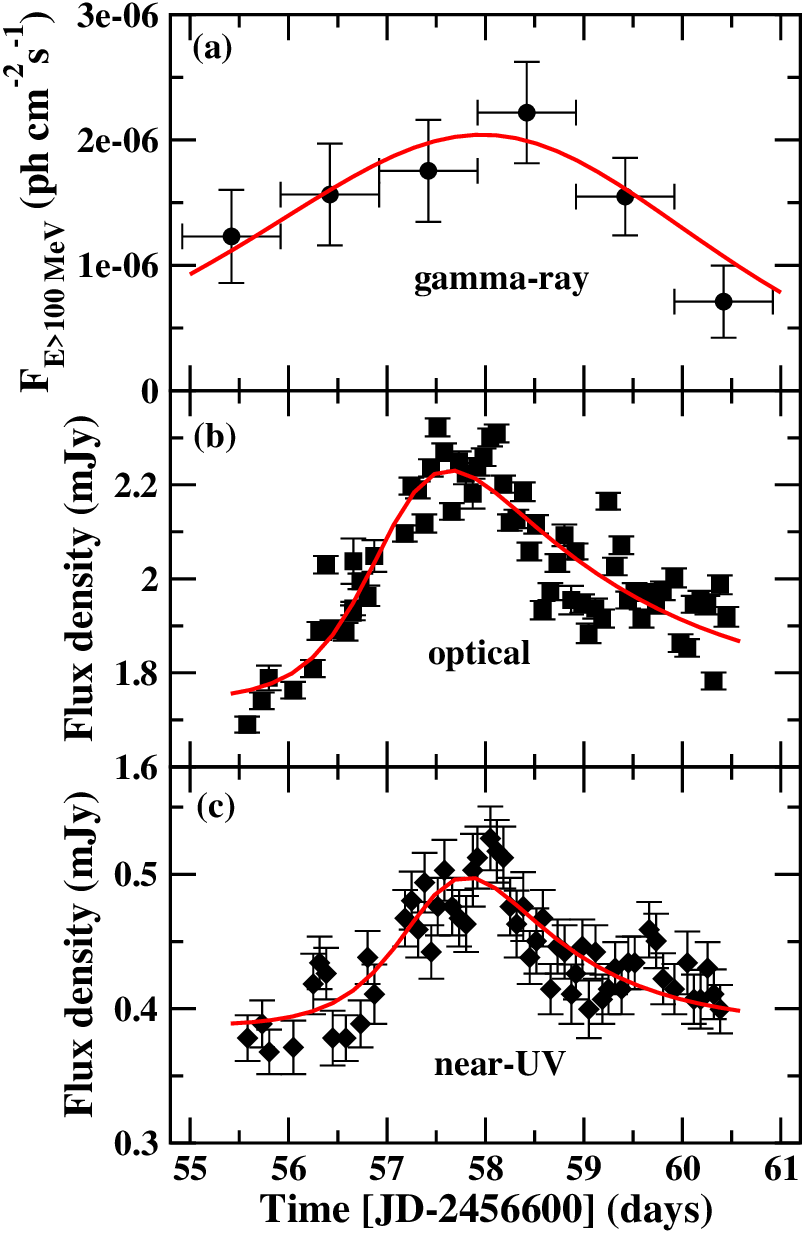}
  \caption{The modeled $\gamma$-ray (a), optical (b), and near-UV (c) flare in 3C~279  using an 
exponential function (see Section \ref{dcf_gamma_opt} for details). The black 
points are the observed data while the red curves represent the fit.   }
\label{plot_fig4b}
\end{figure}

In an alternative approach, we estimated the time corresponding to the peak of the optical/$\gamma$-ray 
flare by modeling their profiles with an exponential function. The flares at optical/$\gamma$-ray 
frequencies can be well approximated with an exponential rise and decay of the form \citep{rani2013_3c273}: 
\begin{equation}
F(t) = 2~F_0 \big [ e^{(t_0 - t)/T_r} + e^{(t - t_0)/T_d} \big ]^{-1},
\end{equation}
where T$_r$ and T$_d$ are the rise and decay times, respectively, and F$_0$ is the source flux at t$_0$ representing
the flare amplitude. The red curves in Fig.~\ref{plot_fig4b} represent the fitted flare components. In Table \ref{tab_flare}, 
we listed the fitted parameters.

Fig.~\ref{plot_fig4b} shows that the rise and decay timescales are very similar for 
the $\gamma$-ray flare, while the optical/near-UV flare has a relatively fast rise and slow decay. Within 
the error bars, the estimated time of the flare peak 
is similar at optical/near-UV and $\gamma$-ray energies, which suggests no time lag between the two. This is 
inconsistent with the DCF analysis results that suggests optical leading $\gamma$-rays by $\sim$0.3~day. The difference is most 
likely due to the faster rise of the optical/near-UV flare, which produces a time lag in the DCF analysis. Given the
relatively mild variations and 1~day time sampling of the $\gamma$-ray light curve, we cannot prove that the optical leads 
the $\gamma$-ray emission. Since we found that the observed time lag seems not to be sensitive to the sampling of the observed 
light curve, however, we cannot ignore the 
possibility that the lag is real. Simultaneous observations at a higher cadence 
would be required to test the hypothesis.  A correlated optical -- $\gamma$-ray flare with optical leading $\gamma$-rays by $\sim$22~hrs was 
previously observed for PKS 1406$-$076 \citep{wagner1995a}. Such correlations seem not to be consistent 
with simple versions of high-energy emissions models, where one expects to see either high-energy flares followed 
by low-energy flares or simultaneous variations.

The simulation only tells us about the significance of correlation given the observational errors. It does not 
account for the possibility that the source has flares at random times in two bands and the times of flares are 
completely uncorrelated. It is therefore essential to compare the long-term behavior of the two light curves. 
We noticed a very similar flaring activity in the long-term optical/$\gamma$-ray light curves; the source 
displayed multiple flares superimposed on top of a broad-outburst observed between  November 2013 and August 
2014. A comparison of the flaring activity suggests a significant correlation between the optical/$\gamma$-ray 
flaring activity,  which supports the idea that the optical/$\gamma$-ray correlation reported here is not a 
random coincidence. 
A detailed analysis of the broadband flaring activity from November 2013 to  August 2014  and its connection 
with the jet kinematics will be given in a separate paper (Rani et al.\ 2016, in preparation).

\subsubsection{X-ray vs.\ optical and gamma-rays} 
We noticed a peak in the X-ray vs.\ optical V passband DCF analysis curve, with a 
 correlation coefficient of 0.53$\pm$0.10 (see Fig.~\ref{plot_fig3}~d), which suggests a possible 
correlation between the two with the former leading the latter with a time lag of $\sim$1.8~days. 
The  significance of the peak is $\sim$90$\%$, however,  below the 95$\%$ confidence level. 
The $\gamma$-ray vs.\ X-ray DCF analysis curve does not show 
any significant peak for the observations (Fig.~\ref{plot_fig3}~a). The DCF analysis therefore does not 
support a correlation between the two.

\subsubsection{Radio vs.\ radio}
Despite having a low fractional variability, variations at different radio bands seem to have similar behavior. 
In the total intensity curves, we noticed only marginal variations, and the same is the case for the EVPA curves. 
A micro-flare is observed in the fractional polarization curve both at 86 and 43~GHz radio bands. 
The observed micro-flare has similar rise ($\sim$1~day) and decay ($\sim$2~days) timescales at the two radio bands, 
suggesting a possible correlation between the two. With only four measurements, 
the radio--radio correlation and also correlation with other bands cannot be tested via the formal DCF 
analysis method.

  \begin{figure}
\includegraphics[scale=0.45,angle=0, trim=0 0 0 0.5, clip]{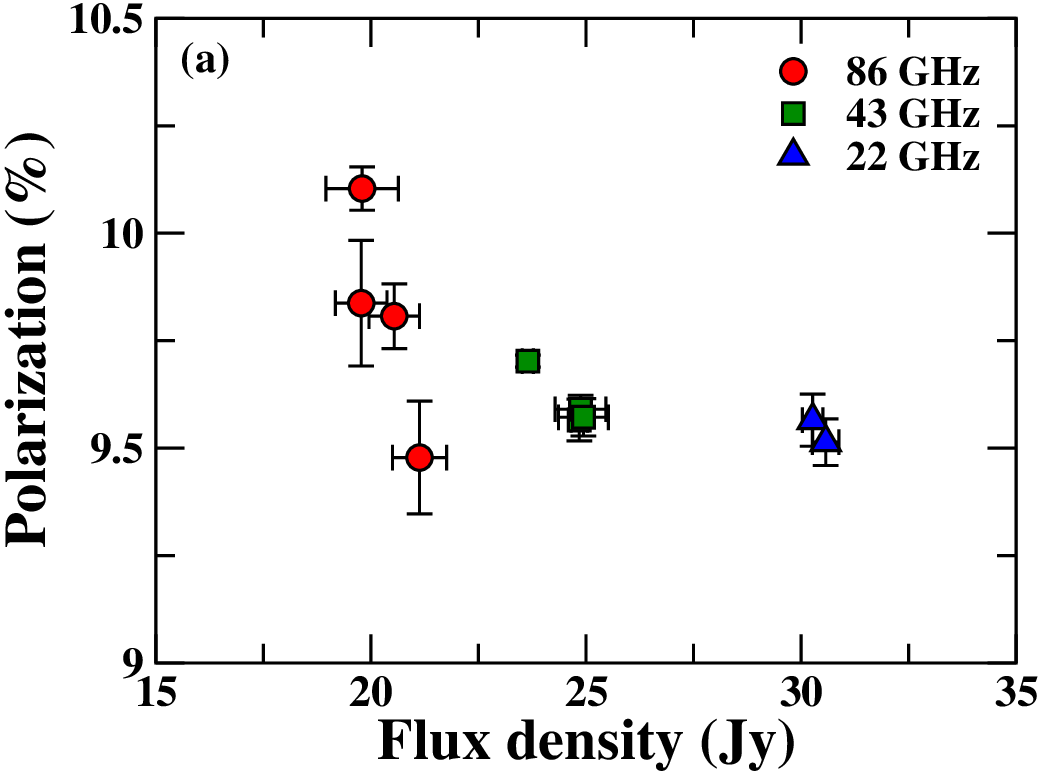}
\includegraphics[scale=0.45,angle=0, trim=0 0 0 0.5, clip]{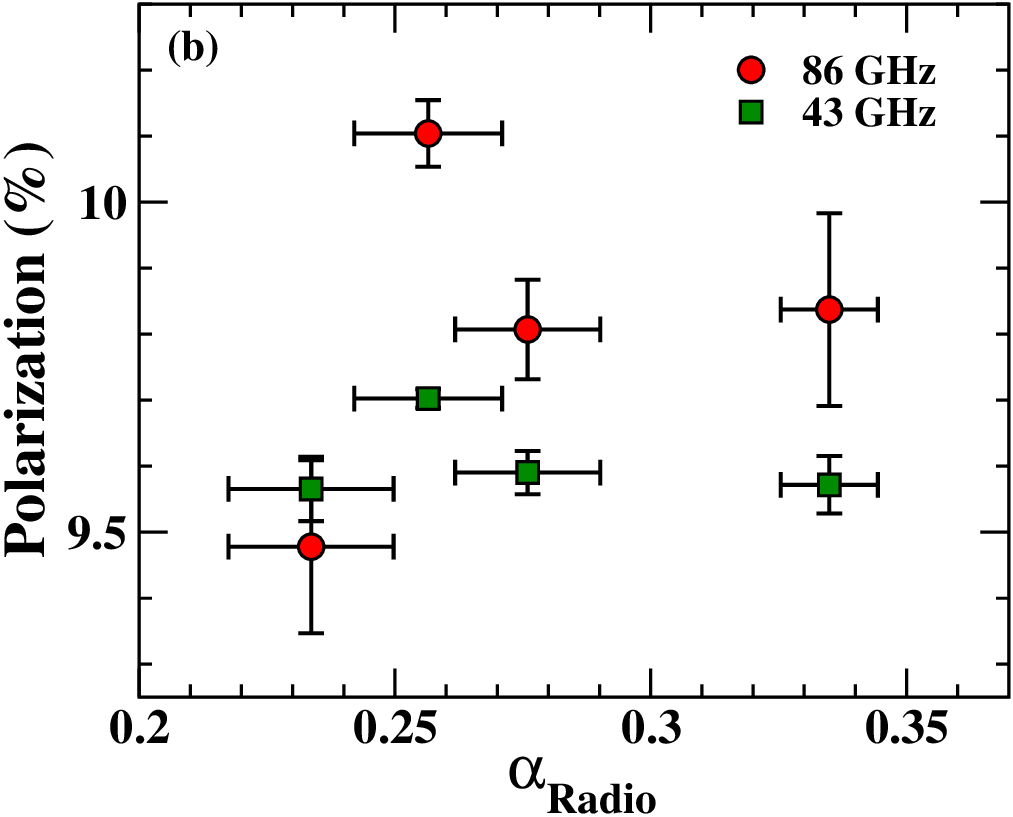}
  \caption{ Fractional linear polarization plotted as a function of 
flux density (a) and spectral index (b) at different radio bands - red circles 
denote 86~GHz band, green squares denote 43~GHz band, 
and blue triangles denote 22~GHz band.       }
\label{plot_fig5}
\end{figure}

\subsection{Flux density vs.\ linear polarization}
Unlike the higher frequencies, the total intensity variations are milder
at radio bands, but we observed a prominent micro-flare in the polarization curve at 
86 and 43~GHz radio bands.  
Figure~\ref{plot_fig5}  shows the fractional linear polarization plotted as a function 
of flux density at different radio bands. 
The degree 
of polarization is higher at higher frequencies. In addition,
 there is an apparent 
anti-correlation between flux density and linear polarization for the source. A formal 
linear Pearson correlation analysis gives r$_p$ = $-$0.64 with a 95.4$\%$ confidence level, 
where r$_p$ is the linear Pearson correlation coefficient. 
A systematic increase in the degree of linear polarization could be expected 
as Faraday depolarization effects decrease with increasing frequency.

Particularly at 86 GHz, a change in polarization degree (PD) occurred while the total intensity and EVPA were relatively 
constant. An increasing PD indicates a more ordered magnetic field or a steepening
of the electron (and also synchrotron) spectrum. The observations also show variations 
in the radio spectrum (see Fig.\ \ref{plot_fig1b}~a), as well as  an indication of steepening of the radio spectrum with an increase in PD (see Fig.\ \ref{plot_fig5}~b); 
however, with only four measurements, we cannot make a statistical claim.   An increase in PD while the spectrum gets softer, although needs to be 
tested further using high cadence observations, provides an hint that  the PD variations could be simply due to 
a spectral change\footnote{The degree of linear 
polarization is equal $\frac{(\alpha +1)}{\alpha + 5/3}$ \citep{rybicki1986}, where $\alpha$ is the spectral 
index.} without any changes in the magnetic field.

\subsection{Faraday Rotation Measure}
\begin{figure}[h]
   \centering
 \includegraphics[scale=0.39,angle=0, trim=0 0 0 0.5, clip]{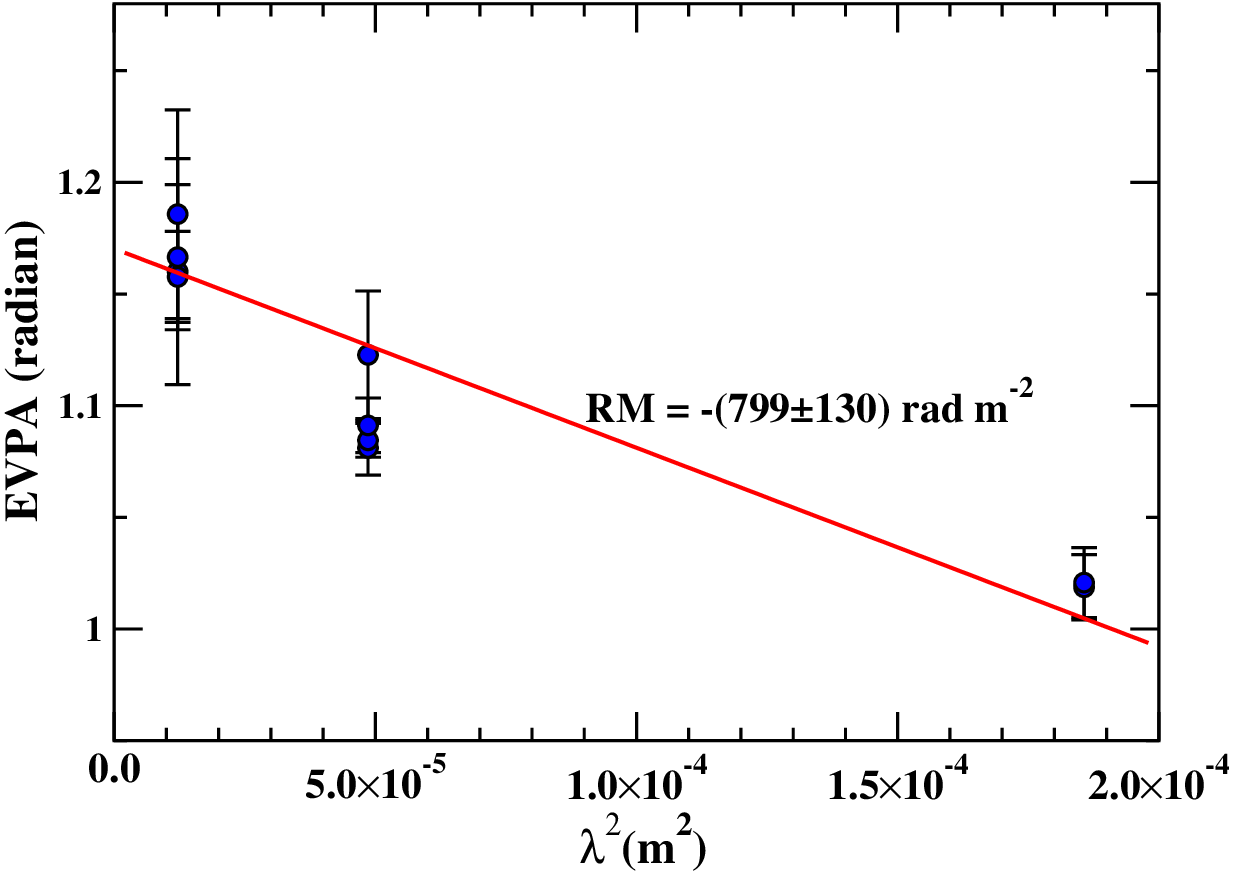}
  \caption{Dependence of EVPA (polarization position angle) on square of wavelength 
in 3C~279. The solid line represents the best fit. }
\label{plot_fig6}
\end{figure}

Figure~\ref{plot_fig1}~(h) shows the EVPA variations measured at 22 (in blue), 43 (in green), 
and 86~GHz (in red) radio bands. The observed difference in the EVPA measurements at different radio bands 
can be used to obtain rotation measure (RM) estimates.  In case of external rotation, the effect can be 
described by a linear dependence between the observed EVPA ($\chi_{obs}$) and wavelength squared ($\lambda^2$)
by 
\begin{eqnarray}
\chi & = & \chi_0 +
 \frac{e^3\lambda^2}{8\pi m_e^2\epsilon_0c}\int N(s)\vec{B}(s)\cdot\,ds \\
 \chi & = & \chi_0 + \textrm{RM}\,\lambda^2
 \end{eqnarray}
\noindent
where $\chi_0$ is the intrinsic polarization angle, $e$ is the electron
charge, $m_e$ is the electron mass, and the integral is taken over the line
of sight from the source to the observer \citep{gabuzda2004}. 
The EVPA values as a function of $\lambda^{2}$ are plotted in Fig.~\ref{plot_fig6}, which 
could be well approximated by a linear fit. We obtained 
EVPA = $-$(799$\pm$130)$\times$$\lambda^{2}$ + 1.16$\pm$0.01. This implies a RM 
of $-$(799$\pm$130)~rad~m$^{-2}$, which is consistent with the previous RM measurements in 3C279  
\citep{zavala2001, lee2015, wardle1998, jorstad2007}. 
Much smaller RM values, 
RM $\sim$ $-$(42 to 74), have also been reported for the source \citep{hovatta2012}. In general, 
 higher RM values are found at higher frequencies, which could be due to an external screen in the 
vicinity of the jet \citep{lee2015, jorstad2007}. 

\section{Discussion and Conclusions}

Extremely bright GeV/TeV flares at minutes to hours timescales are a common characteristic of  
$\gamma$-ray-bright blazars. The rapid flares usually do not have a low-frequency  
counterpart \citep{rani2013,rani2015, abdo2010c, hayashida2012}, 
making it difficult to interpret the radiation processes and acceleration mechanisms involved.  
We organized a multi-wavelength campaign to understand the physical processes responsible 
for the origin of extreme $\gamma$-ray flaring activity observed in 3C~279 in December 2013. 
The source was observed from radio (including polarization) to $\gamma$-rays.

A relatively mild $\gamma$-ray flare was observed in 3C~279 during our campaign period. Compared to 
the extremely bright and rapid flares ($t_{var} \sim$few hours) seen at other times, this flare had
relatively longer variability timescales ($t_{rise}$  and $t_{decay} \sim2.5$~days).     
Nevertheless, the $\gamma$-ray flare was accompanied by a prominent flare at 
optical and near-UV passbands, and we found a significant correlation between the two. 
The optical/near-UV flare, however, had a relatively fast rise ($\sim$0.4~day) and slow decay 
($\sim$1.5~days). The flux variations (both at optical and $\gamma$-ray frequencies) were accompanied by no 
significant spectral variations.  
Broadband spectral modeling of the flaring activity by \citet{hayashida2015} for a period that has 
3 days overlap with our observations suggests a substantial contribution 
of external-Compton radiation from the broad-line region. Their analysis also indicates that the location 
of the $\gamma$-ray emission region is comparable with the broad-line region radius.  
The observed optical--$\gamma$-ray correlation suggests a co-spatiality of the two emission regions.

 We observed a slight hint of variability (significance $\sim$1.4~$\sigma$) in the 0.3--10~keV X-ray 
light curve. The formal cross-correlation analysis suggests no correlation between the X-ray 
and $\gamma$-ray variations. There is a hint of possible 
correlation between the X-ray and the optical variations with the former leading the latter by $\sim$1.8~days; however, the 
confidence level of the correlation is below 95$\%$. A micro-flare peaking in the fractional polarization curve 
at 43 and 86~GHz radio bands coincides with the first peak (X1) in the X-ray light curve. 
Observations at a better cadence would be needed to test this apparent correlation. 
Unlike the higher frequencies, the total intensity variations are not pronounced 
at radio bands. We noticed a micro-flare in polarization degree (PD), while the variations in EVPA curves 
are not statistically significant. An indication of spectral steepening at radio bands 
with an increase in PD suggests that the PD variations could be due to a spectral change 
without  any change in the magnetic field (B-field).

The presence of contemporaneous optical/$\gamma$-ray flaring activity is not always the case for blazars. 
However,  studies over the past few years indicate a similarity in the variability properties of 
the two bands \citep{cohen2014, chatterjee2012}, suggesting the same population of electrons being responsible 
for the synchrotron (optical) and inverse-Compton ($\gamma$-rays) radiation. For 3C~279, \citet{hayashida2012} 
reported a good correlation between the optical and $\gamma$-ray flares in 2008--2012. However later 
in 2013--2014, this correlation seemed to be less obvious \citep{hayashida2015}. A comparison of 
long-term optical/$\gamma$-ray flaring activity between November 2013 and
August 2014 including our campaign period suggests a significant correlation between the two 
(Rani et al.\ 2016, in preparation). The observed 
optical--X-ray correlations in 3C~279 also present a complicated picture. A detailed analysis of the 
optical--X-ray correlations between 1996 and 2007 by \citet{chatterjee2008} suggests that for some flares 
optical lead X-rays while for others it is vice-versa. Additionally, an orphan X-ray flare was observed in 
the source in 2008--2010 
\citep{abdo2010c, hayashida2012}. Presence of multiple emission regions \citep{marscher2014, hayashida2015}, 
different dependencies of magnetic field density and energy density of the external radiation field 
\citep{janiak2012}, and/or particle acceleration in a stratified jet \citep{rani2013,rani2015} have all 
been suggested as plausible scenarios.

During our campaign period, a short-duration  ($\sim$5~day) flare at optical/near-UV frequencies is found to be 
significantly correlated with a contemporaneous flare at $\gamma$-rays. The $F_{var}$ at  $\gamma$-ray does not exceed that at optical/near-UV bands, 
which is usually not  what had been observed for 3C~279. The $F_{var}$ at $\gamma$-rays 
is usually much higher than or at least comparable to  that at optical frequencies \citep[][Rani et al.\ 2016, in preparation]
{hayashida2012, hayashida2015}. 
The optical flare has a marginally significant correlation with the X-ray variations. 
Nearly simultaneous spectral analysis \citep{hayashida2015} suggests superposition of two 
spectral components, as a single-zone leptonic model failed to explain the X-ray spectrum. 
A coincidence of a micro-flare in the fraction of polarization at radio bands with the X-ray flare certainly 
 emphasizes its jet base origin.

In a tentative scenario, a single population of electrons seems to be responsible for the optical/near-UV 
and $\gamma$-ray flares. X-rays could be produced via synchrotron self-Compton with seed photons 
coming from  short-mm radio bands as is supported by the coincidence of the increase of radio polarization  (at 43 and 86~GHz bands) 
with the X-ray flux. The X-ray flare leading the optical flare could be expected if the high-energy 
electrons take longer to accelerate compared to the lower-energy electrons \citep{boettcher2007}. 
 As a 
consequence, 
the radio and X-ray flares lead the optical/$\gamma$-ray flares. 
An alternative explanation could be that the particles are accelerated 
in a different region. The presence of multiple sub-components in the emission could also be possible.

In leptonic models, the relative 
amplitude variations of synchrotron and IC flares could be simply determined by the B-field, the 
number of emitting electrons ($N_e$), and the Doppler factor ($\delta$) \citep{chatterjee2008}. A spectral 
hardening at radio frequencies with an increase in polarization degree does not support 
the B-field variations. An increase in both $N_e$ and $\delta$ causes a larger enhancement in the IC-flux 
compared to the synchrotron flux \citep{chatterjee2008}. As a result,  one expects to see larger variability 
amplitude for the IC flare compared to the synchrotron flares; however, for our observations, the reverse 
is true. This could be 
because of different dependencies of B-field energy density and 
external radiation field energy density on the distance along the jet from the central engine \citep{janiak2012}. 
It is important to note that we consider leptonic models for our interpretation; however, hadronic models 
are equally able to reproduce a similar kind of variability.

\section{Acknowledgements}
The {\it Fermi}/LAT Collaboration acknowledges the generous support of a number of agencies
and institutes that have supported the {\it Fermi}/LAT Collaboration. These include the National
Aeronautics and Space Administration and the Department of Energy in the United States, the
Commissariat \`a l'Energie Atomique and the Centre National de la Recherche Scientifique / Institut
National de Physique Nucl\'eaire et de Physique des Particules in France, the Agenzia Spaziale
Italiana and the Istituto Nazionale di Fisica Nucleare in Italy, the Ministry of Education,
Culture, Sports, Science and Technology (MEXT), High Energy Accelerator Research Organization
(KEK) and Japan Aerospace Exploration Agency (JAXA) in Japan, and the K.\ A.\ Wallenberg
Foundation, the Swedish Research Council and the Swedish National Space Board in Sweden.
Additional support for science analysis during the operations phase is gratefully acknowledged 
from the Istituto Nazionale di Astrofisica in Italy and the Centre National d'\'Etudes Spatiales 
in France. This research was supported by an appointment to the NASA Postdoctoral Program
at the Goddard Space Flight Center, administered by Universities Space Research Association 
through a contract with NASA.
 We would like to thank the referee for his/her constructive comments.
BR acknowledges the help of M. B{\"o}ttcher, Vassilis Karamanavis, Greg Madejski, Roopesh Ojha, Jeremy Perkins, 
and Dave Thompson for fruitful discussions 
and comments that improved the manuscript. BR is thankful to D. Emmanoulopoulos for the 
useful discussions on statistical analysis. 
The KVN is a facility operated by the Korea Astronomy and Space Science Institute. The KVN operations 
are supported by KREONET (Korea Research Environment Open NETwork) which is managed and operated 
by KISTI (Korea Institute of Science and Technology Information). 
KVS is supported by the RFBR grant 14-02-31789. 
SSL and KS are supported by the National Research Foundation of Korea (NRF) grant funded by the Korea 
government(MSIP) (No.\ NRF-2016R1C1B2006697).


\end{document}